\documentclass[preprint]{revtex4}%
\usepackage{graphicx}%
\usepackage{amsmath}%
\usepackage{amsfonts}%
\usepackage{amssymb}

\def\s{{\sigma}}
\def\e{{\epsilon}}
\def\k{{ {\bf k} }}

\def\Q{{ {\bf Q} }}

\def\w{{\omega}}
\def\a{{\alpha}}
\def\b{{\beta}}

\def\g{{\gamma}}

\begin{document}
\title{
Impurity-induced in-gap state and $T_{\rm c}$ \\
in sign-reversing $s$-wave superconductors:\\
analysis of iron oxypnictide superconductors
}
\author{Yuko \textsc{Senga}$^{1,2}$ and Hiroshi \textsc{Kontani}$^{1,2}$}
\date{\today }

\begin{abstract}
The sign-reversing fully gapped superconducting state,
which is expected to be realized in oxypnictide superconductors,
can be prominently affected by nonmagnetic impurities
due to the interband scattering of Cooper pairs.
We study this problem based on the isotropic two-band BCS model:
In oxypnictide superconductors, 
the interband impurity scattering $I'$ is not equal to the intraband one $I$.
In the Born scattering regime, the reduction in $T_{\rm c}$ is sizable 
and the impurity-induced density of states (DOS) is prominent if $I\sim I'$, 
due to the interband scattering.
Although impurity-induced DOS can yield a
power-law temperature dependence in $1/T_1$, 
a sizable suppression in $T_{\rm c}$ is inevitably accompanied.
In the unitary scattering regime, in contrast, impurity effect is very small 
for both $T_{\rm c}$ and DOS except at $I=I'$.
By comparing theory and experiments, we expect that the degree of anisotropy 
in the $s_\pm$-wave gap function strongly depends on compounds.
\end{abstract}
%\draft

\address{
$^1$ Department of Physics, Nagoya University, 
Furo-cho, Nagoya 464-8602, Japan. 
\\
$^2$ 
JST, Transformative Research-Project on Iron Pnictides (TRIP), 
Chiyoda, Tokyo 102-0075, Japan.
%JST, TRIP, Nagoya University, Furo-cho, Nagoya 464-8602, Japan. 
}
 
%74.20.-z, 74.20.Fg, 74.20.Rp

\sloppy

\maketitle

%%%%%%%%%%%%%%%%%%
%Introduction
%%%%%%%%%%%%%%%%%%
\section{Introduction}

Recently, the mechanism of superconductivity in 
high-$T_{\rm c}$ superconductors with FeAs layers 
 \cite{Hosono,Ren,XHChen,GFChen,Eisaki}
has been attracting considerable attentions.
The superconducting state is realized by introducing carrier
into the parent compound, which shows the spin density wave (SDW) 
state at $T_N\sim 130$K \cite{Luetkens,Liu}.
In the SDW state, the ordered magnetic moment is $\sim0.3 \ \mu_{\rm B}$
and the ordering vector is $\Q\approx (\pi,0)$
 \cite{Moss,Luetkens,neutron-SDW,neutron-noRP}.
NMR studies had clearly shown that 
the singlet superconducting state is realized in iron oxypnictides
 \cite{Kawabata1,Kawabata2,Zheng}.
A fully gapped superconducting state has been determined
by the penetration depth measurement \cite{Matsuda},
angle-resolved photoemission spectroscopy (ARPES)
 \cite{ARPES1,ARPES2,ARPES3,ARPES4}, specific heat measurement \cite{Mu},
and so on.

In the first-principle band calculations \cite{Singh,Ishibashi,Arita}, 
the Fermi surfaces in iron oxypnictides are composed of
two hole-like Fermi pockets around the $\Gamma=(0,0)$ point 
and  two electron-like Fermi pockets around M$=(\pi,0),(0,\pi)$ points.
The nesting between the hole and electron pockets
is expected to give rise to the SDW state in undoped compounds.
In doped compounds without SDW order, 
the antiferromagnetic (AF) fluctuations with $\Q\approx (\pi,0)$
is expected to induce a fully gapped $s$-wave state with sign reversal,
which is called the $s_\pm$-wave state
 \cite{Kuroki,Mazin,RG,Nomura,Yanagi,Ikeda,Tesanovic,Fuseya}.
Moreover, near the SDW boundary, the Hall coefficient
and Nernst signal show prominent anomalous behaviors
 \cite{Liu,Nernst,Kawabata2},
which are similar to those observed in high-$T_{\rm c}$ cuprates
and in Ce$M$In$_5$ ($M$=Co,Rh,Ir) \cite{Nakajima}.
Theoretically, these anomalous transport phenomena 
indicate the existence of strong AF fluctuations
 \cite{Kontani-review}.
At the same time, huge residual resistivity far beyond the
$s$-wave unitary scattering is expected to appear near the SDW boundary
theoretically \cite{Impurity}.

To investigate the pairing symmetry of superconductivity,
impurity effects on the superconducting state offer us decisive informations.
In iron oxypnictide superconductors, 
impurity effect on $T_{\rm c}$ due to Co, Ni, or Zn 
substitution for Fe sites is very small or absent
 \cite{Kawabata1,Kawabata2,Co,Ni,Zn}.
This result clearly rules out the possibility of line-node superconductivity.
One may also expect that the $s$-wave state with sign reversal
is also eliminated, since the Cooper pair is destroyed by
the interband scattering induced by impurities.
However, we have recently shown that $T_{\rm c}$ is almost unchanged
by strong (unitary) impurities, 
since the interband impurity scattering potential $I'$ is
different from the intraband one $I$ \cite{Senga}.
The reason for this unexpected result is that the effective interband 
scattering is renormalized to zero in the unitary limit except at $I=I'$
in the $T$-matrix approximation.
Therefore, the experimental absence of impurity effect on $T_{\rm c}$ 
in iron oxypnictides is well understood in terms of the $s_\pm$-wave state.
On the other hand, $T_{\rm c}$ will be prominently reduced by
short-range weak (Born) impurities \cite{Senga}.

Recently, several authors had revealed that 
in-gap density of states (DOS) is induced by impurities
in the $s_\pm$-wave state using the Born approximation 
for general value of $x\equiv |I'/I|$ \cite{Chubukov}, or using
the $T$-matrix approximation only for $x=1$ \cite{Parker,Bang}.
They also demonstrated that 
the relation $1/T_1\propto T^3$ under $T_{\rm c}$,
which had been reported by several groups \cite{Zheng,Ishida,Mukuda,Grafe},
can be reproduced by the impurity-induced DOS.
However, the assumed impurity parameters ($n_{\rm imp}$, $I$ and $I'$)
also yields a sizable suppression in $T_{\rm c}$ 
according to the analysis in Ref. \cite{Senga}.
Furthermore, impurity-induced DOS should be sensitive to the value of 
$x$ in the unitary scattering regime, as suggested in ref. \cite{Senga}.
Therefore, we have to study the impurity effects on the DOS and $T_{\rm c}$
for general $x$, and compare their relationships in detail.

In this paper, we investigate the 
impurity-induced DOS and $T_{\rm c}$ in the $s_\pm$-wave state 
using the $T$-matrix approximation for general $x$.
We stress that $x$ is not unity in iron oxypnictides since hole and electron 
pockets are not composed of the same $d$-orbitals.
In the Born or intermediate scattering regime,
a sizable impurity-induced DOS appears for $x\gtrsim0.7$, and therefore
$1/T_1$ may deviate from a simple exponential behavior.
Although impurity-induced DOS can yield a power-law temperature dependence 
in $1/T_1$ \cite{Chubukov,Parker,Bang},
we find that a sizable suppression in $T_{\rm c}$ is inevitably accompanied.
The anisotropy in the $s_\pm$-wave superconducting gap,
which had been predicted theoretically 
\cite{Kuroki,RG},
might be responsible for the power-law temperature dependence of 
$1/T_1$ under $T_{\rm c}$ as discussed in ref. \cite{Nagai}.
In contrast, unitary impurities affect both the superconducting DOS 
and $T_{\rm c}$ only slightly, except at $I=I'$.

\section{$T$-matrix approximation in the two-band BCS model}
\label{sec:T-matrix}
As studied in refs. \cite{Chubukov,Bang,Senga,Tesanovic},
the $s_\pm$-wave state is realized in the two-band BCS model
if we introduce the interband repulsive interaction, which represents 
the AF fluctuations due to the interband nesting in iron oxypnictides.
In the present paper, 
we study the impurity effect using the $T$-matrix approximation 
for general $I'/I$.
In the presence of mass enhancement due to many-body effect, $m^*/m_0>1$,
both the superconducting gap and the impurity effect
(or impurity concentration $n_{\rm imp}$)
are renormalized by the factor $(m^*/m_0)^{-1}$.
In the present analysis, we neglect the mass-enhancement for simplicity.

In the Nambu representation, the two-band BCS model 
%without impurity
is given by \cite{Schrieffer,Allen}
\begin{eqnarray}
{\hat H}^0=\sum_\k {\hat c}_\k^\dagger {\hat H}_\k^0 {\hat c}_\k ,
\end{eqnarray}
where ${\hat c}_\k^\dagger= (c_{\k\uparrow}^{\a\dagger}, 
c_{\k\uparrow}^{\b\dagger},c_{-\k\downarrow}{\a},c_{-\k\downarrow}^{\b})$, and 
\begin{eqnarray}
{\hat H}_\k^0=\left(
\begin{array}{cccc}
\e_\k^\a & 0 & \Delta_\a & 0 \\
0 & \e_\k^\b & 0 & \Delta_\b \\
\Delta_\a & 0 & -\e_\k^\a & 0 \\
0 & \Delta_\b & 0 & -\e_\k^\b \\
\end{array}
\right) .
 \label{eqn:H0}
\end{eqnarray}
In eq. (\ref{eqn:H0}), $\e_\k^\a, \e_\k^\b$ are the band dispersions
measured from the Fermi level.
Since we consider the isotropic $s_\pm$ superconducting state,
only the DOSs for both bands at the Fermi level ($N_\a,N_\b$)
are taken into consideration in the present BCS study.
$\Delta_\a, \Delta_\b$ in eq. (\ref{eqn:H0})
are the superconducting gap.
When only the inter-band repulsive interaction ($g_{\a\b}=g_{\b\a}>0$)
is taken into consideration, the gap equation without impurities 
is given as \cite{Parker,Bang,Senga},
\begin{eqnarray}
\Delta_{\a(\b)}= -g_{\a\b} N_{\b(\a)} T\sum_n 
f_{\b(\a)}(i\e_n)\theta(\e_n-|\w_c|),
 \label{eqn:Gapeq}
\end{eqnarray}
where $\e_n=\pi T(2n+1)$ is the fermion Matsubara frequency,
and $\w_c$ is the cutoff energy.
$N_{\b(\a)}$ is the DOS for $\b(\a)$-band at the Fermi energy 
in the normal state per spin.
$f_{\b(\a)}(\e)$ is the local anomalous Green function
for $\b(\a)$ band, which will be given later.
Since $f_{\b(\a)}\propto\Delta_{\b(\a)}$, the $s_\pm$-state 
$\Delta_\a=-\Delta_\b$ is realized for $g_{\a\b}>0$ \cite{Parker,Bang,Senga}.
Moreover, $|\Delta_\a/\Delta_\b| \sim (N_\b/N_\a)^{1/2}$
since $f_\a/f_\b \sim \Delta_\a/\Delta_\b$.

The Nambu matrix representation for the impurity potential is given as
\begin{eqnarray}
{\hat I}=\left(
\begin{array}{cccc}
I & I' & 0 & 0 \\
I' & I & 0 & 0 \\
0 & 0 & -I & -I' \\
0 & 0 & -I' & -I \\
\end{array}
\right) .
\end{eqnarray}
We can assume that $I,I'\ge0$ without losing generality.
In the presence of impurities, 
the Green function in the Nambu representation is given by \cite{Allen}
\begin{eqnarray}
{\hat G}_\k({\tilde \w})=({\tilde \w}{\hat 1}-{\hat H}_\k^0-{\hat \Sigma}({\tilde \w}))^{-1} ,
\end{eqnarray}
where ${\tilde \w}\equiv \w+i\delta$ ($\delta=+0$), and 
${\hat \Sigma}({\tilde \w})$ is the self-energy due to impurities.

Hereafter, we derive ${\hat \Sigma}({\tilde \w})$ in the 
%self-consistent 
$T$-matrix approximation, which gives the exact result for 
$n_{\rm imp}\ll1$ for any strength of $I, I'$.
The $T$-matrix in the Nambu representation is given by
\begin{eqnarray}
{\hat T}({\tilde \w})=({\hat 1}-{\hat I}\cdot{\hat g}({\tilde \w}))^{-1}{\hat I} ,
 \label{eqn:T}
\end{eqnarray}
where ${\hat g}({\tilde \w})\equiv \frac1N \sum_\k {\hat G}_\k({\tilde \w})$ 
is the local Green function, which is given by \cite{Allen}
\begin{eqnarray}
{\hat g}({\tilde \w})=\left(
\begin{array}{cccc}
g_\a({\tilde \w}) & 0 & f_\a({\tilde \w}) & 0 \\
0 & g_\b({\tilde \w}) & 0 & f_\b({\tilde \w}) \\
f_\a({\tilde \w}) & 0 & g_\a({\tilde \w}) & 0 \\
0 & f_\b({\tilde \w}) & 0 & g_\b({\tilde \w}) .\\
\end{array}
\right)
 \label{eqn:local-g}
\end{eqnarray}
In the above expression,
$g_i$ and $f_i$ ($i=\a,\b$) are given by
\begin{eqnarray}
g_i({\tilde \w})&=& -\pi N_i \frac{{\tilde \w} Z_i({\tilde \w})}
 {\sqrt{-({\tilde \w} Z_i({\tilde \w}))^2+(\Delta_i+\Sigma_i^a({\tilde \w}))}} ,
 \label{eqn:g}\\
f_i({\tilde \w})&=& -\pi N_i \frac{\Delta_i+\Sigma_i^a({\tilde \w})}
 {\sqrt{-({\tilde \w} Z_i({\tilde \w}))^2+(\Delta_i+\Sigma_i^a({\tilde \w}))}} ,
 \\
Z_i({\tilde \w})&=& 1-\frac1{2{\tilde \w}}(\Sigma_i^n({\tilde \w})-\Sigma_i^n(-{\tilde \w})) ,
\end{eqnarray}
where $\Sigma_i^n$ and $\Sigma_i^a$ ($i=\a,\b$) are the 
normal and anomalous self-energies, respectively.
In the $T$-matrix approximation, 
the self-energies are given by using eq. (\ref{eqn:T}) as
\begin{eqnarray}
\Sigma_\a^n({\tilde \w})&=& n_{\rm imp} T_{11}({\tilde \w}), \ 
\Sigma_\b^n({\tilde \w})= n_{\rm imp} T_{22}({\tilde \w}) , \\
\Sigma_\a^n(-{\tilde \w})&=& -n_{\rm imp} T_{33}({\tilde \w}), \ 
\Sigma_\b^n(-{\tilde \w})= -n_{\rm imp} T_{44}({\tilde \w}) , \\
\Sigma_\a^a({\tilde \w})&=& n_{\rm imp} T_{13}({\tilde \w}), \ 
\Sigma_\b^a({\tilde \w})= n_{\rm imp} T_{24}({\tilde \w}) .
 \label{eqn:S24}
\end{eqnarray}

In the fully self-consistent $T$-matrix approximation, we have to solve 
eqs. (\ref{eqn:Gapeq}) and (\ref{eqn:local-g})-(\ref{eqn:S24}) 
self-consistently.
In this paper, however, we solve only 
eqs. (\ref{eqn:local-g})-(\ref{eqn:S24}) self-consistently,
by neglecting the impurity effect on $\Delta_\a$ and $\Delta_\b$ 
in eq. (\ref{eqn:Gapeq}).
This approximation is justified when 
the reduction in $\Delta_{\a(\b)}$ due to impurity pair-breaking is small.

\section{Numerical results}

Here, we discuss the impurity effect on the DOS and $T_{\rm c}$
in the $s_\pm$-wave superconducting state,
based on the numerical results given by the 
%self-consistent
$T$-matrix approximation.
%In the two-band BCS model with repulsive interband interaction,
%$|\Delta_\a/\Delta_\b|=(N_\b/N_\a)^{1/2}$ is satisfied \cite{Senga}.

%%%%%%%%%%%%%%%%%%%%%%%%%%%%%%%%%
\begin{figure}[!htb]
{\hspace{8mm}} {\bf (a)} {\hspace{78mm}}  {\bf (b)}\\
\includegraphics[width=.49\linewidth]{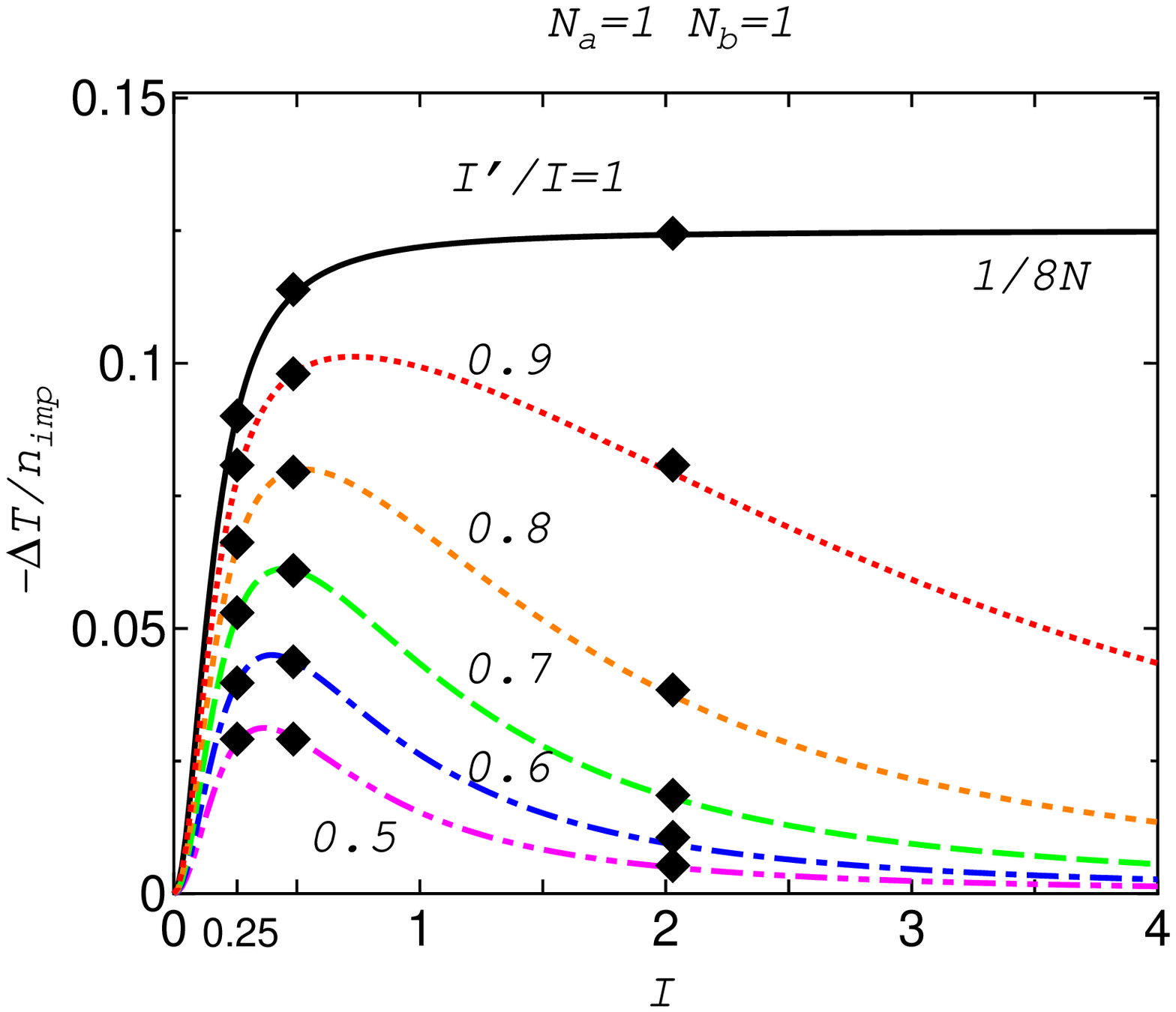}
\includegraphics[width=.49\linewidth]{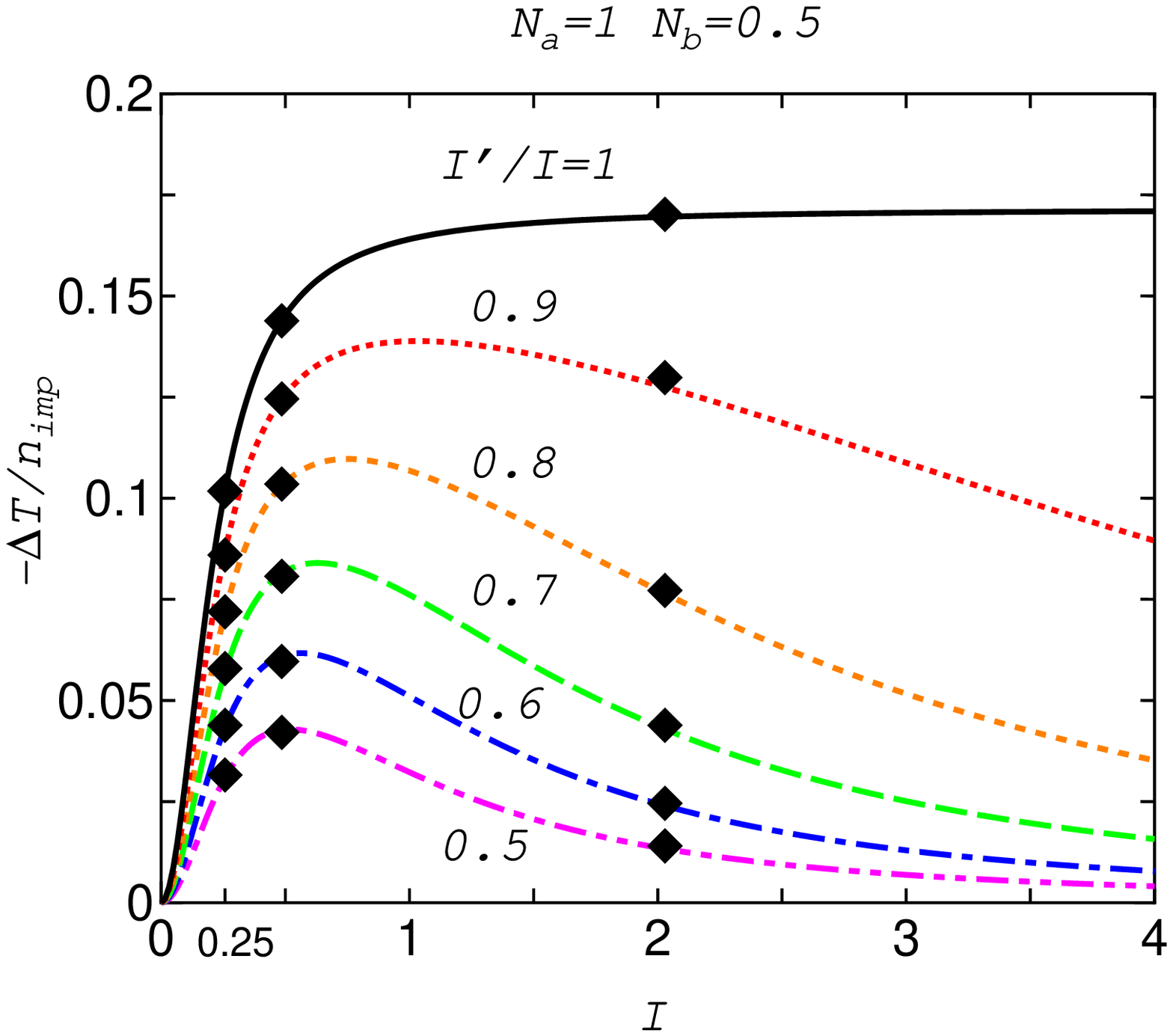}
\caption{
$-\Delta T_{\rm c}/n_{\rm imp}$ as a function of $I$
given by the $T$-matrix approximation
in the case of (a) $N_\a=N_\b=1$ and (b) $N_\a=1, N_\b=0.5$.
In both figures, the unit of energy is $1/N_\a$,
which corresponds to 18000 K for (a) and 14000 K for (b),
since $N_\a+N_\b=1.31$ eV$^{-1}$ in iron oxypnictides.
$-\Delta T_{\rm c}/n_{\rm imp}$ for ($x=1$, $I=\infty$)
is $1/8N_\a \sim 2300$ K for (a), and $1/5.84N_\a \sim 2400$ K for (b).
The superconducting DOS for parameters denoted by filled diamonds are
shown in Figs. \ref{fig:DOS1} and \ref{fig:DOS2}.
}
\label{fig:Tc1}
\end{figure}
%%%%%%%%%%%%%%%%%%%%%%%%%%%%%%%%%

\subsection{Impurity effect on $T_{\rm c}$}
As derived in Ref. \cite{Senga}, the expression for 
the reduction in $T_{\rm c}$ per impurity concentration
based on the two-band BCS model is given as
\begin{eqnarray}
-\frac{\Delta T_{\rm c}}{n_{\rm imp}}= 
\frac{\pi^2\left[ \ 3(N_\a+N_\b)-2\sqrt{N_\a N_\b} \ \right]{I'}^2}
{8{\bar A}} ,
\label{eqn:DTc}
\end{eqnarray}
For $n_{\rm imp}\ll1$, the transition temperature is given by
$T_{\rm c}= T_{\rm c}^0 -(-\Delta T_{\rm c}/n_{\rm imp})\cdot n_{\rm imp}$,
where $T_{\rm c}^0$ is the transition temperature without impurities.
In eq. (\ref{eqn:DTc}), 
${\bar A}= 1+\pi^2I^2(N_\a^2+N_\b^2)+2N_\a N_\b\pi^2{I'}^2
 + N_\a^2N_\b^2\pi^4(I^2-{I'}^2)^2$.
In the case of $x\equiv I'/I =1$,
the right hand side of eq. (\ref{eqn:DTc}) is 
$[3(N_\a+N_\b)-2\sqrt{N_\a N_\b}]/[8(N_\a+N_\b)^2]+O(I^{-2})$
in the unitary regime.
In the case of $x \ne 1$, in contrast,
eq. (\ref{eqn:DTc}) is given by
$x^2[3(N_\a+N_\b)-2\sqrt{N_\a N_\b}]/[8\pi^2 N_\a^2 N_\b^2(1-x^2)I^2]
 +O(I^{-4})$.
Therefore, eq. (\ref{eqn:DTc}) approaches zero in the case of 
$x\ne1$ in the unitary regime.

Figure \ref{fig:Tc1} (a) shows $-\Delta T_{\rm c}/n_{\rm imp}$
given in eq. (\ref{eqn:DTc}) in the case of $N_\a=N_\b=1$.
In iron oxypnictides, the total DOS per Fe atom 
($N_\a+N_\b$) is 1.31 eV$^{-1}$ per spin \cite{Singh}.
Then, $1/N=1$ corresponds to 18000 K.
When $x=1$, $-\Delta T_{\rm c}/n_{\rm imp}$
approaches $1/8N\sim 2300$ K in the unitary regime ($IN \gg 1$).
Therefore, the superconductivity in iron oxypnictides will vanish 
only at $n_{\rm imp}\approx 8N\cdot T_{\rm c}^0= 0.01\sim0.02$ [$1\sim2$ \%].
When $x \ne1$, in high contrast, 
$-\Delta T_{\rm c}/n_{\rm imp}$ decreases and approaches zero
as $I$ increases in the unitary regime,
since the effective interband scattering is renormalized as
$I'_{\rm eff} \sim I'\cdot(IN)^{-2} \ll I'$ \cite{Senga}.

According to the first principle calculations,
$N_\b/N_\a\gtrsim 0.7$ in iron oxypnictides \cite{NaNb}.
Here, we study the case of $N_\b/N_\a=0.5$ in order to clarify the 
the impurity effect on $T_{\rm c}$ for the 
the particle-hole asymmetric case; $N_\b/N_\a\ne1$.
Figure \ref{fig:Tc1} (b) shows $-\Delta T_{\rm c}/n_{\rm imp}$
for $N_\a=1$ and $N_\b=0.5$.
According to eq. (\ref{eqn:DTc}),
$-\Delta T_{\rm c}/n_{\rm imp}$ for $x=1$ and $I=\infty$
is $1/5.84N_\a \sim 2400$ K, by taking account of the relation
$N_\a+N_\b=1.31$ eV$^{-1}$ in iron oxypnictides.
By comparing with the results for $N_\a=N_\b=1$ in Fig. \ref{fig:Tc1} (a),
we find that $-\Delta T_{\rm c}/n_{\rm imp}$ is 
insensitive to the value of $N_\b/N_\a$,
under the condition that $N_\a+N_\b=$constant.

%%%%%%%%%%%%%%%%%%%%%%%%%%%%%%%%%
\begin{figure}[!htb]
\includegraphics[width=.55\linewidth]{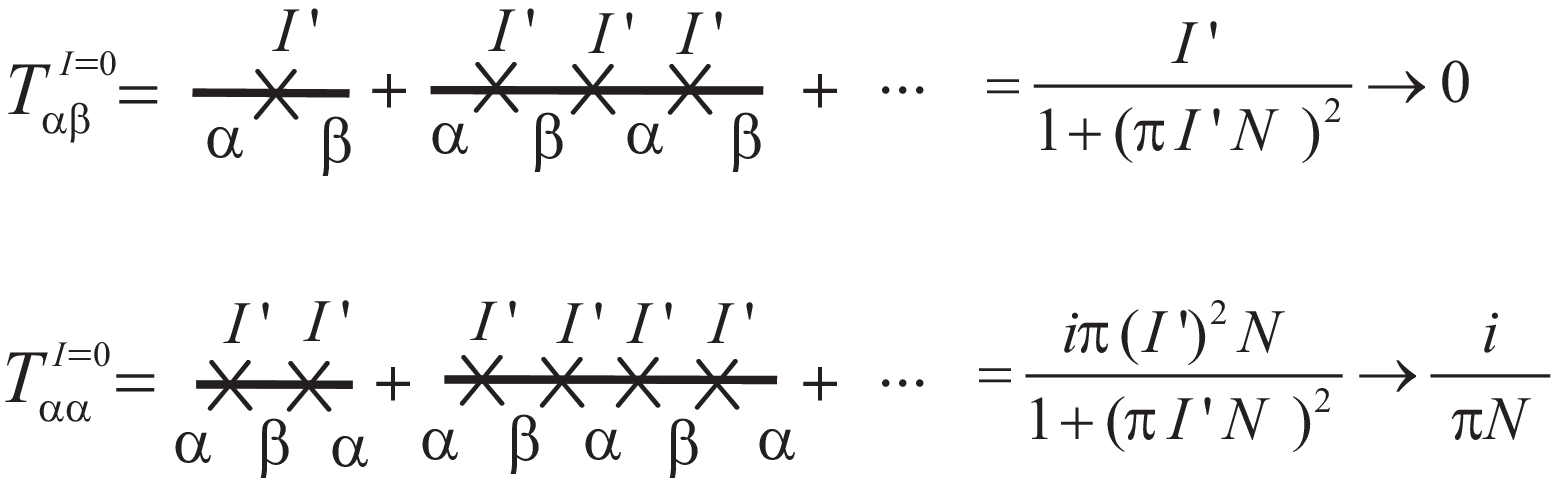}
\caption{
Intraband and interband
$T$-matrices in the normal state, $T_{\a\a}^{I=0}$ and $T_{\a\b}^{I=0}$.
}
\label{fig:T-matrix}
\end{figure}
%%%%%%%%%%%%%%%%%%%%%%%%%%%%%%%%%
Previously, impurity effect on $T_{\rm c}$ in two-band BCS models
had been studied by many authors in various contexts
 \cite{Preosti,Kulic,Ohashi,Mg1,Mg2,Mg3}, and
it was found that $T_{\rm c}$ is unchanged in the unitary limit
 \cite{Kulic,Ohashi}.
However, eq. (\ref{eqn:DTc}) for $s_\pm$-state had not been derived.
Here, we present a clear explanation why the 
interband scattering (pair breaking) is absent in the unitary regime,
which had not been discussed previously.
Figure \ref{fig:T-matrix} shows the intraband and interband
$T$-matrices in the normal state, $T_{\a\a}^{I=0}$ and $T_{\a\b}^{I=0}$,
in the case of $I=0$ and $N_\a=N_\b=N$.
Apparently, $T_{\a\b}^{I=0}$ approaches zero for $I'\rightarrow\infty$.
Next, we consider $T_{\a\b}$ for general ($I,I'$).
If we construct $T_{\a\b}$ of ($T_{\a\a}^{I=0}$, $T_{\a\b}^{I=0}$, $I$),
it contains at least one ${\hat T}_{\a\b}^{I=0}$.
For this reason,
interband $T$-matrix is expected to approach zero in the unitary regime.
This expectation is correct unless $x=1$, as shown in Ref. \cite{Senga}.

%%%%%%%%%%%%%%%%%%%%%%%%%%%%%%%%%
\begin{figure}[!htb]
{\hspace{8mm}} {\bf (a)} {\hspace{78mm}}  {\bf (b)}\\
\includegraphics[width=.49\linewidth]{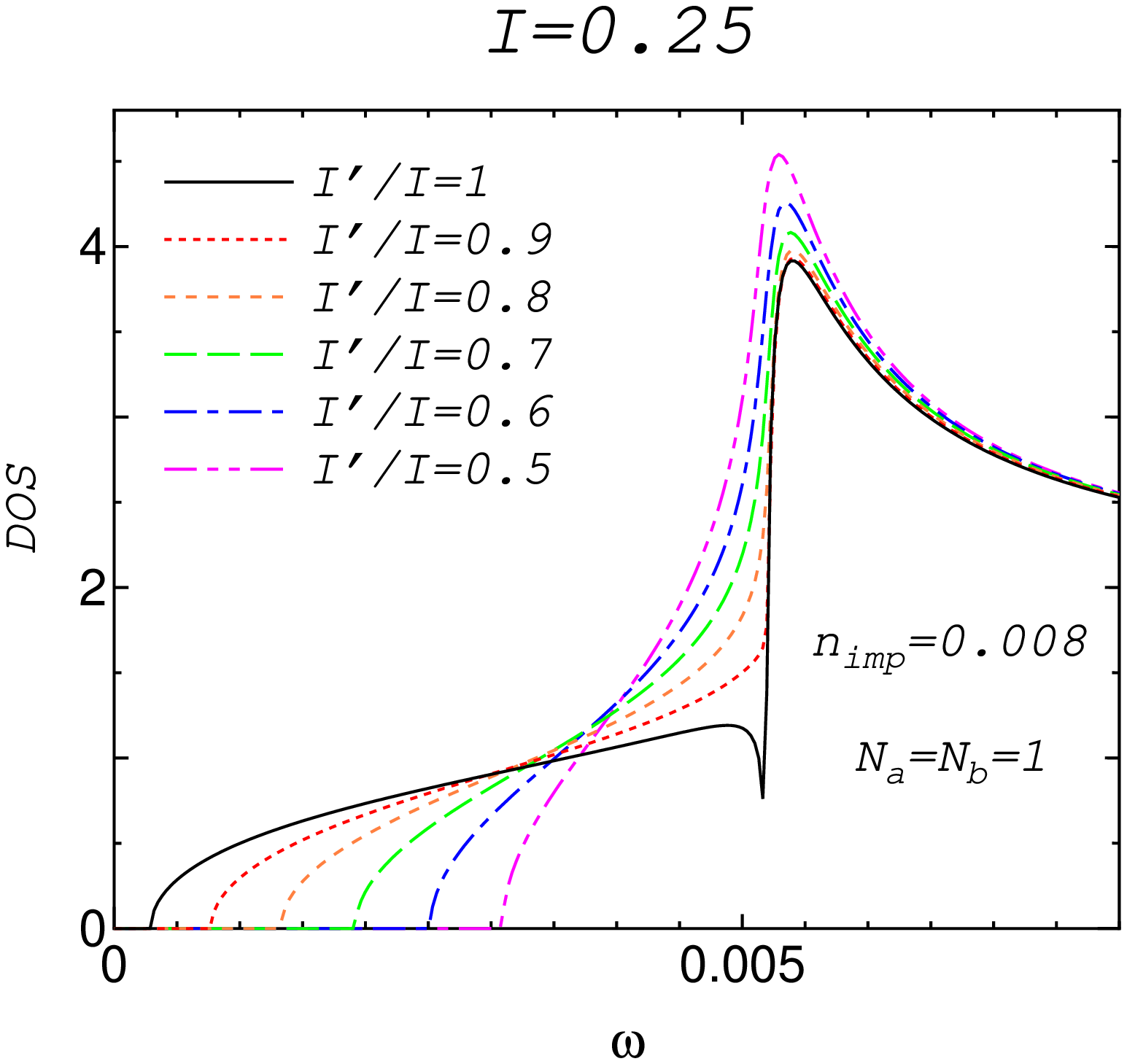}
\includegraphics[width=.49\linewidth]{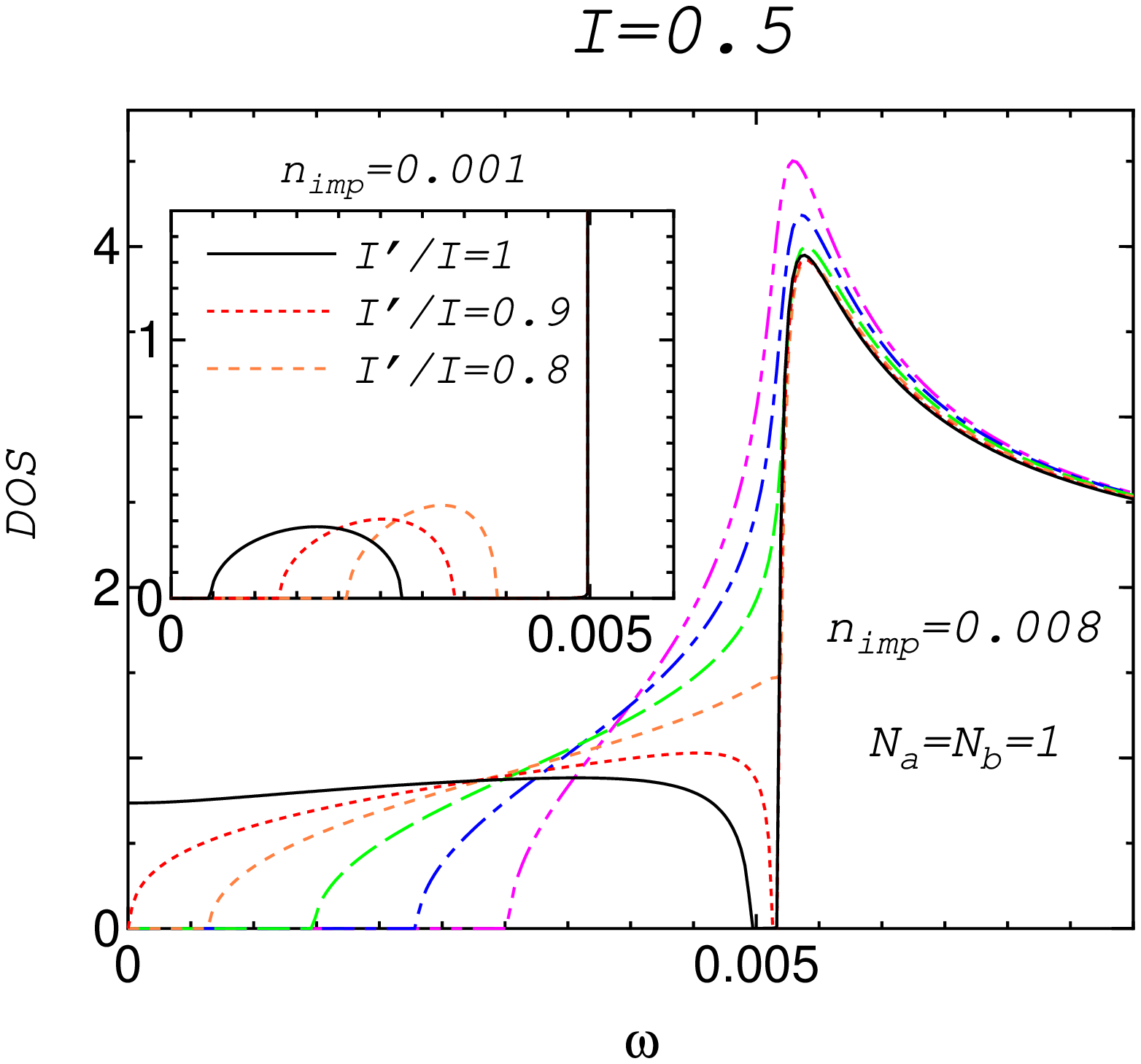}
{\hspace{38mm}} {\bf (c)} {\hspace{78mm}}  {\bf (d)}\\
\includegraphics[width=.49\linewidth]{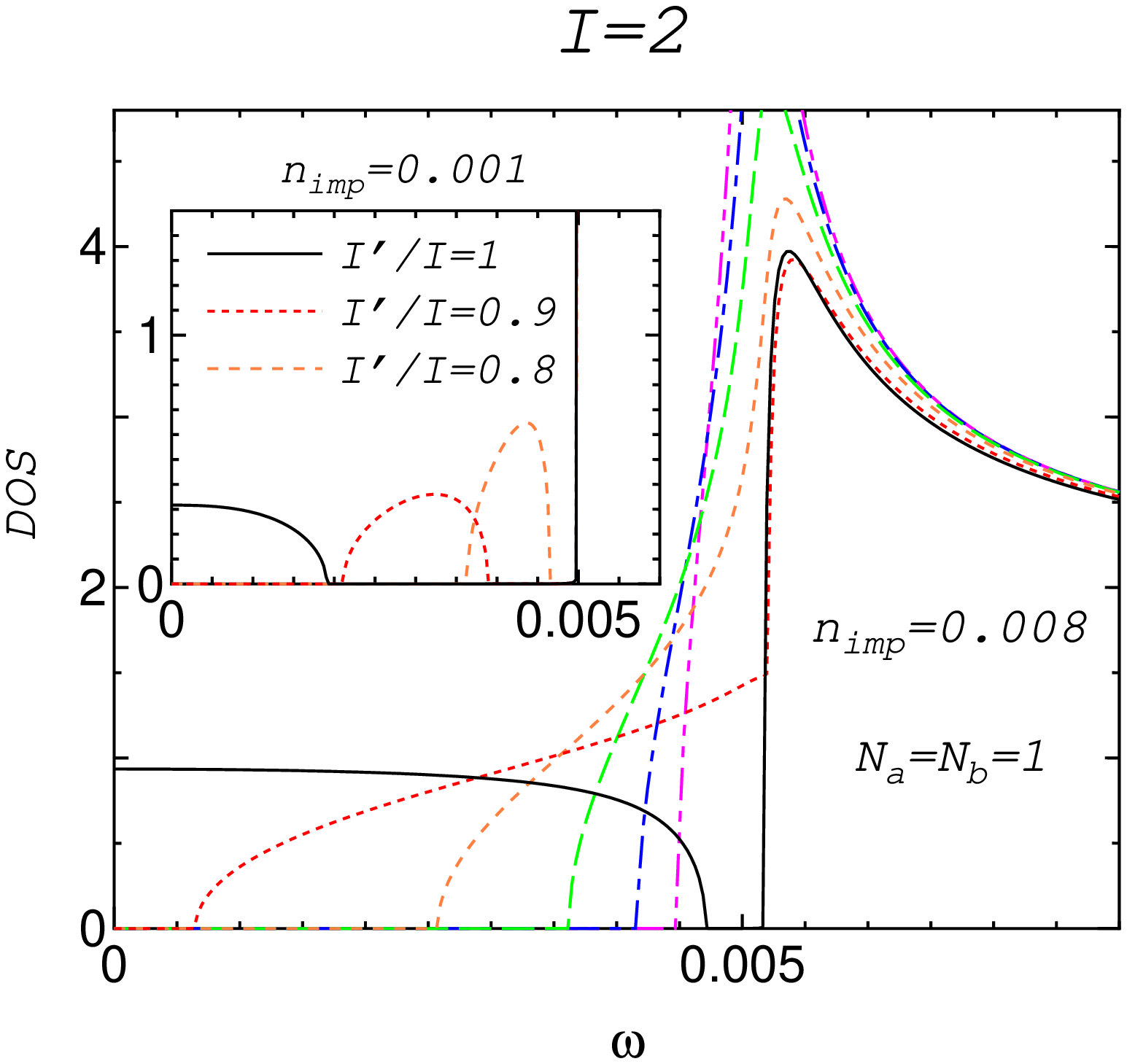}
\includegraphics[width=.49\linewidth]{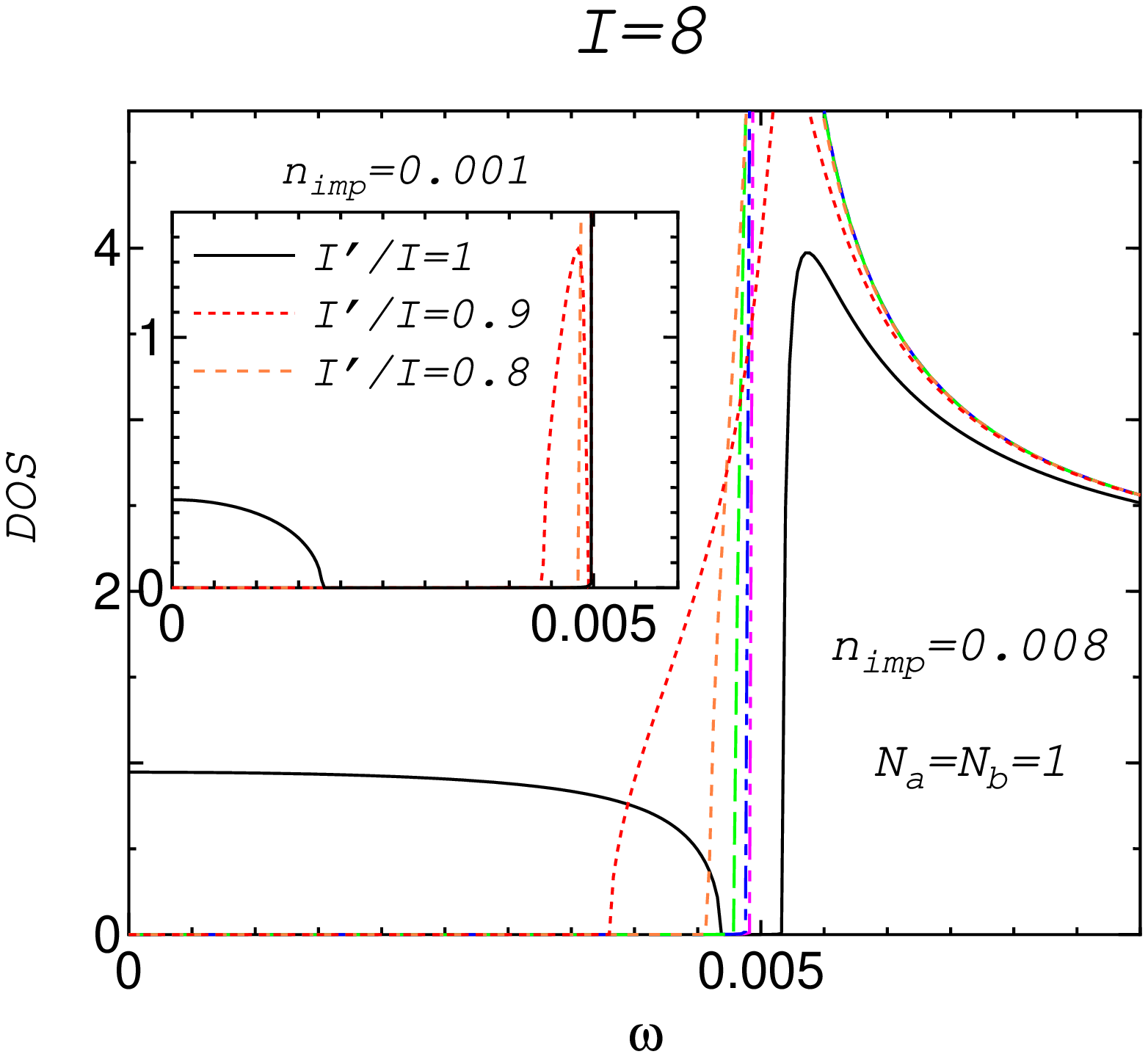}
\caption{
Obtained DOS in the superconducting state for 
$N_\a=N_\b=1 \ (\sim 0.66 \ {\rm eV}^{-1})$
and $\Delta_\a=-\Delta_\b=0.005\ (\sim 100 \ {\rm K})$, 
in the case of (a) $I=0.25$, (b) $I=0.5$, (c) $I=2$, and (d) $I=8$.
Impurity concentration $n_{\rm imp}$ is $0.008$.
The insets in (b)-(d) present the DOS for $n_{\rm imp}=0.001$.
Note that $N(-\w)=N(\w)$.
}
\label{fig:DOS1}
\end{figure}
%%%%%%%%%%%%%%%%%%%%%%%%%%%%%%%%%

\subsection{Impurity effect on the DOS}
In the $s_\pm$-wave superconducting state, impurity interband scattering 
not only reduces $T_{\rm c}$, but also induces the in-gap state
in the superconducting DOS
 \cite{Preosti,Ohashi,Parker,Bang2}.
The DOS is given by the imaginary part of the local Green function,
which is expressed in eq. (\ref{eqn:g}), as follows:
\begin{eqnarray}
N(\w)= -\frac1\pi {\rm Im}\{ g_\a({\tilde \w})+g_\b({\tilde \w}) \} .
\end{eqnarray}
If $n_{\rm imp}\ll1$,
the obtained DOS will be reliable for any $I$ and $I'$  
in the present 
%self-consistent 
$T$-matrix approximation.
Figure \ref{fig:DOS1} shows the DOS in the superconducting state 
in the case of $N_\a=N_\b=1$ and $\Delta_\a=-\Delta_\b=0.005$
for $n_{\rm imp}=0.008$.
$|\Delta_{\a,\b}|=0.005$ corresponds to 90 K.
Experimentally, in Ba$_{0.6}$K$_{0.4}$Fe$_2$As$_2$,
$|\Delta|=11\sim12$ meV for $\a$ Fermi surface (hole-like)
and for $\g$ and $\delta$ Fermi surfaces (electron-like),
and $|\Delta|=5.8$ meV for $\b$ Fermi surface (hole-like) \cite{ARPES4}.
As shown in Fig. \ref{fig:DOS1} (a), the superconducting gap is almost 
filled by the impurity-induced DOS when $I=I'=0.25$ (Born regime),
which corresponds to $\sim 5000$ K.
In this case,
the nuclear relaxation ratio $1/T_1$ shows a 
power-law temperature dependence
since the impurity-induced DOS is approximately linear in $\w$,
like in line-node superconductors.

\begin{figure}[!htb]
\includegraphics[width=.5\linewidth]{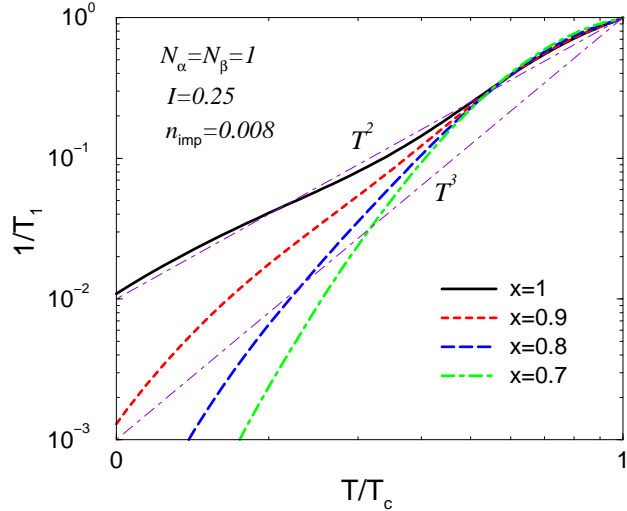}
\caption{
Obtained $1/T_1$ below $T_{\rm c}$ for $I=0.25$ and $x=0.7\sim1.0$.
For $x\le0.8$, $1/T_1$ decreases much faster than $T^3$ at low temperatures
because of the absence of the in-gap state near $\w=0$.
Since the quasiparticle damping rate is 
$\gamma= n_{\rm imp}\pi N(I^2+I'^2)/{\bar A}$ \cite{Senga},
$\gamma/\Delta=0.18$ for $x=1$.
}
\label{fig:T1T}
\end{figure}

Figure \ref{fig:T1T} shows $1/T_1$ below $T_{\rm c}$ for $I=I'=0.25$,
where $\sigma\equiv (\pi N I)^2/(1+(\pi N I)^2)=0.38$. 
We put $|\Delta_{\a(\b)}|= 0.005\sqrt{1-T/T_{\rm c}}$ and $T_{\rm c}=0.002$.
The method of calculation is explained in Refs. \cite{Parker,Bang2}.
For $x=1.0$ and 0.9, $1/T_1$ is inside of $T^2$- and $T^3$-lines
for $T_{\rm c}>T>0.1T_{\rm c}$,
consistently with the analysis by Parker et al. for $\s=0.4$ \cite{Parker}.
In these cases, however, reduction in $T_{\rm c}$ due to impurities, 
which is given by $n_{\rm imp}$ times $-\Delta T_{\rm c}/n_{\rm imp}$ 
in Fig. \ref{fig:Tc1} (a), reaches $13$ K.
%As $x\equiv|I'/I|$ decreases from unity, $-\Delta T_{\rm c}/n_{\rm imp}$
%gradually decreases, and the impurity-induced DOS moves to the gap edge
%at the same time.
The estimated reduction in $T_{\rm c}$ would be {\it underestimated}
since $-\Delta T_{\rm c}/n_{\rm imp}$ is an increase function of $n_{\rm imp}$
for $T_{\rm c}\lesssim T_{\rm c}^0/2$ \cite{Bang,Allen}.
In all cases we have studied (Fig. \ref{fig:DOS1} (a)-(d)),
power-law behavior in $1/T_1$ for $T\ll T_{\rm c}$
due to the galpess superconducting state 
always accompanies a sizable suppression in $T_{\rm c}$,
$-\Delta T_{\rm c} \gtrsim 10$ K, for $|\Delta|= 90$ K.
When $|\Delta|= 40$ K, the gapless superconducting state 
can be realized when $-\Delta T_{\rm c} \sim 6$ K.
Thus, it will be difficult to ascribe the experimental relation
$1/T_1\propto T^3$ below $T_{\rm c}$ \cite{Ishida,Grafe,Mukuda} 
in clean samples with high $T_{\rm c}$ to the impurity effect.
% in the isotropic $s_\pm$-wave state.

%%%%%%%%%%%%%%%%%%%%%%%%%%%%%%%%%
%%%%%%%%%%%%%%%%%%%%%%%%%%%%%%%%%

In the intermediate ($I=0.5$) or unitary ($I\ge2$) regime,
in Figs. \ref{fig:DOS1} (b)-(d),
a large impurity-induced DOS appears at the zero energy in the case of $x=1$,
which is consistent with previous theoretical studies \cite{Parker,Bang2}.
%This result is since $f_\a+f_\b=0$.
In this case, however, $-\Delta T_{\rm c}\sim 13$ K for $n_{\rm imp}=0.008$.
If we put $x \le0.9$, in contrast, $-\Delta T_{\rm c}$ in the unitary regime
is prominently reduced as shown in Fig. \ref{fig:Tc1}.
At the same time, the impurity-induced DOS quickly moves to the gap edge
and disappears, as demonstrated in Figs. \ref{fig:DOS1} (c) and (d).
The reason for these results is that the effective interband scattering 
is renormalized as $I'_{\rm eff} \sim I'\cdot(IN)^{-2} \ll I'$ 
in the unitary regime \cite{Senga}.

%%%%%%%%%%%%%%%%%%%%%%%%%%%%%%%%%
\begin{figure}[!htb]
{\hspace{8mm}} {\bf (a)} {\hspace{78mm}}  {\bf (b)}\\
\includegraphics[width=.49\linewidth]{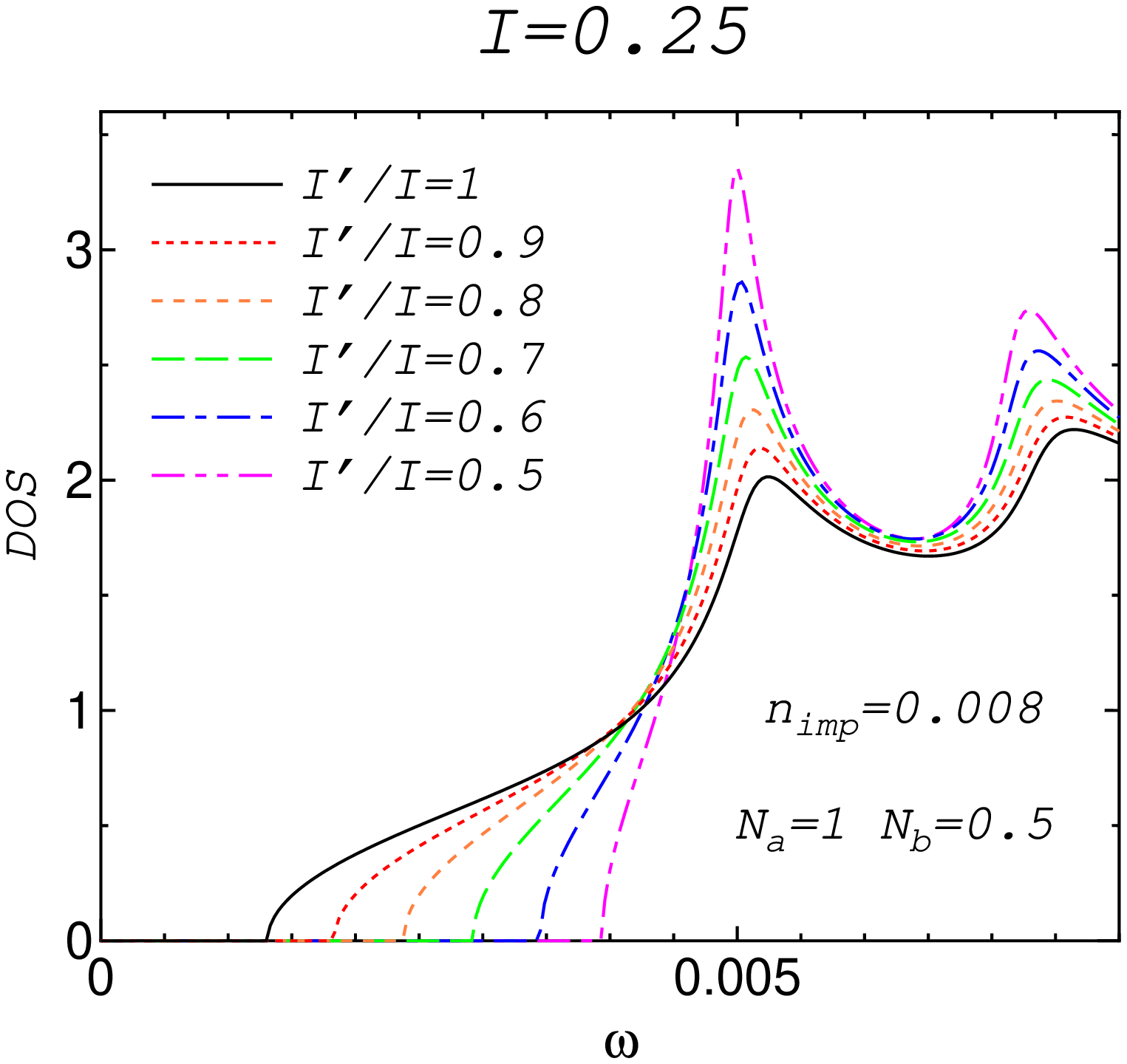}
\includegraphics[width=.49\linewidth]{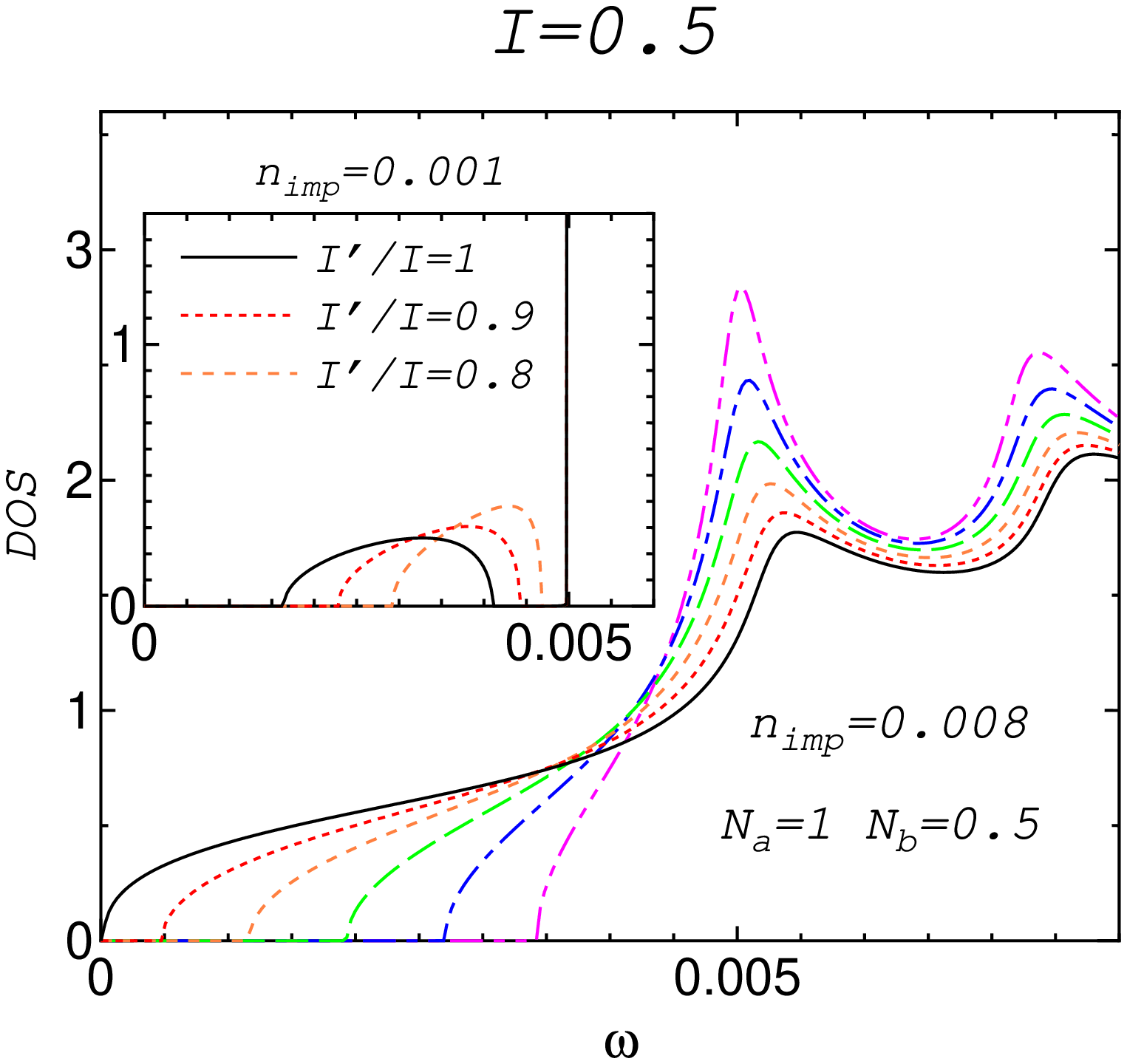}
{\hspace{8mm}} {\bf (c)} {\hspace{78mm}}  {\bf (d)}\\
\includegraphics[width=.49\linewidth]{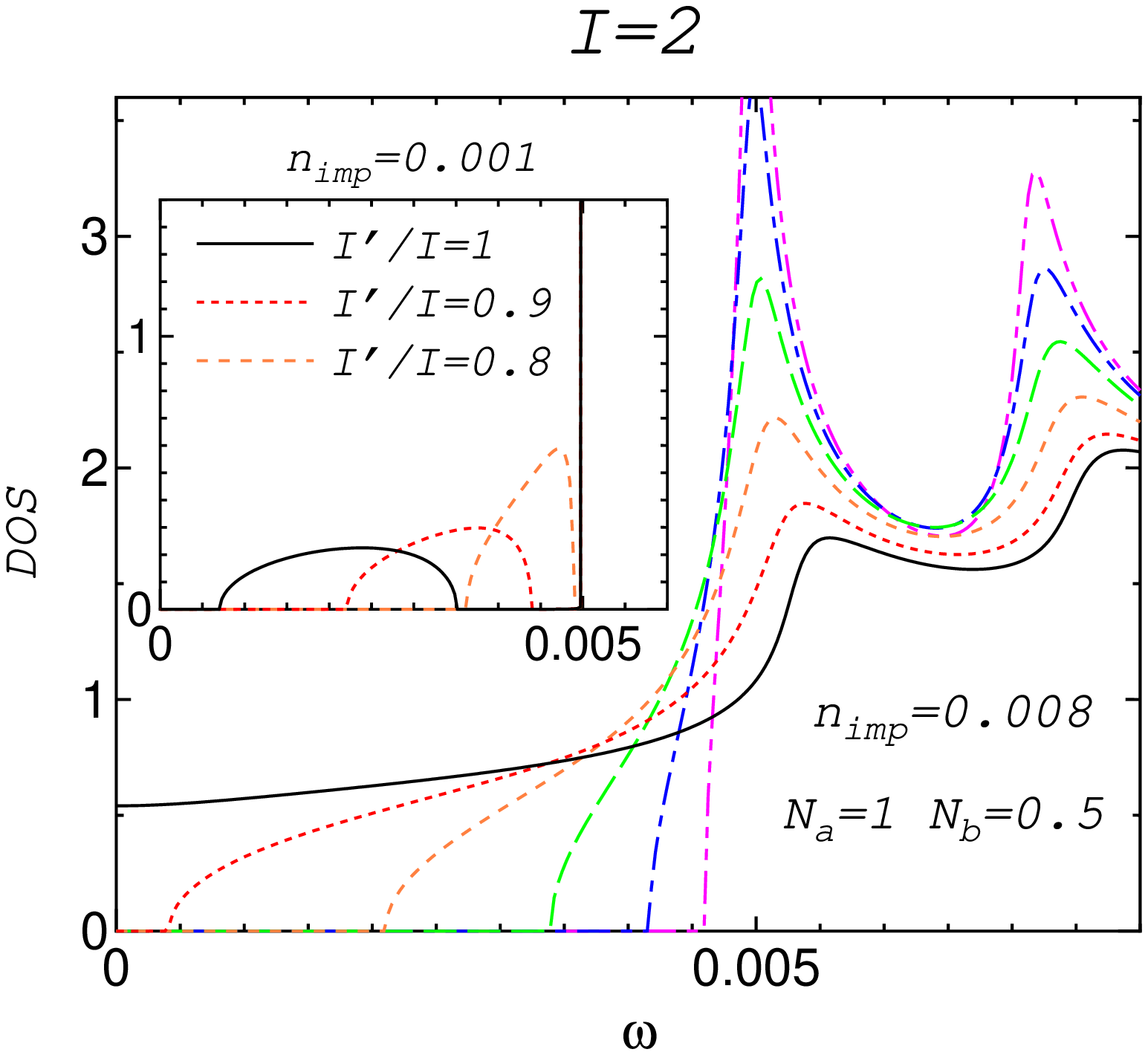}
\includegraphics[width=.49\linewidth]{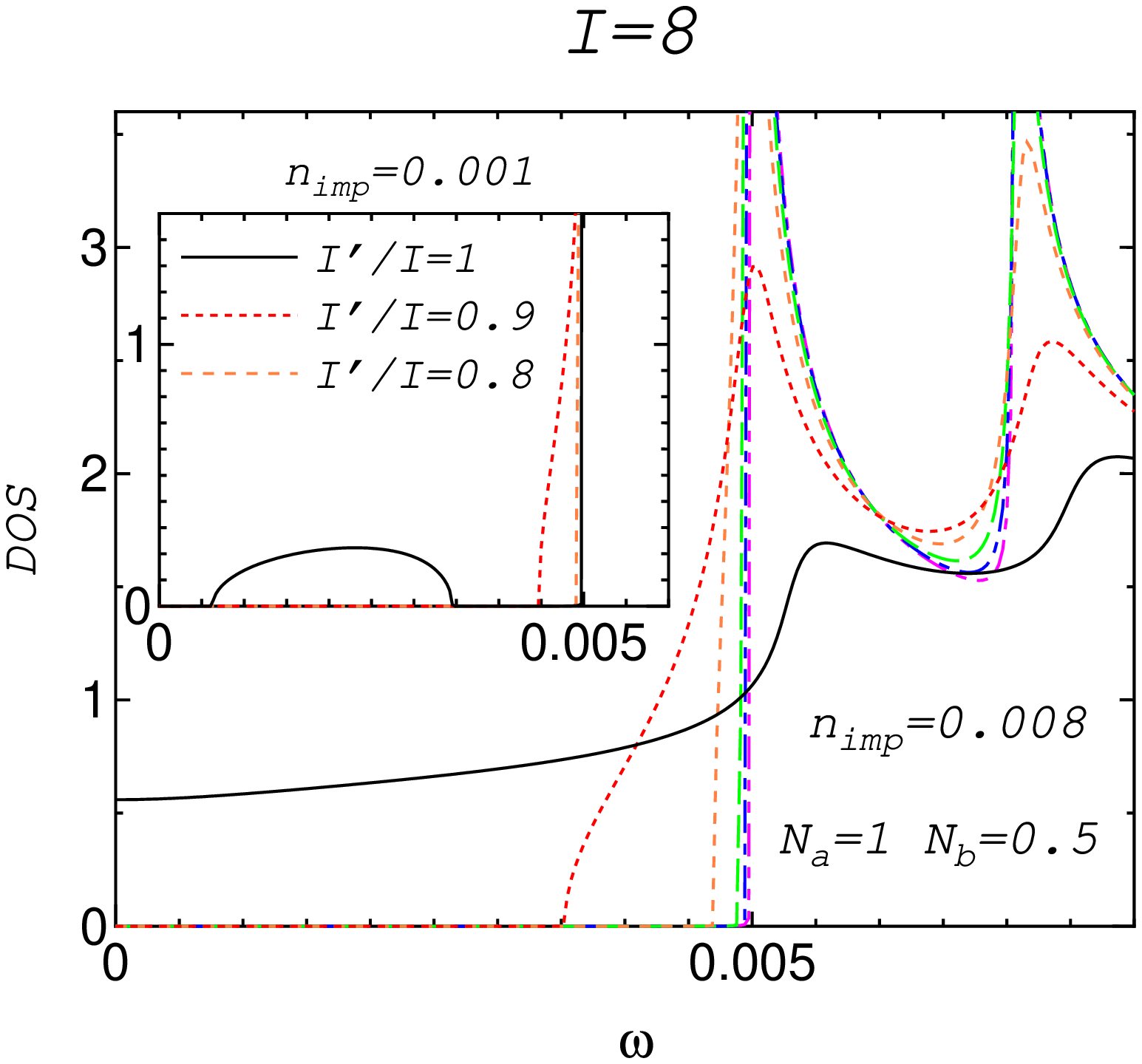}
\caption{
Obtained DOS in the superconducting state for 
$N_\a=1, N_\b=0.5$ and $\Delta_\a=0.005, \Delta_\b=-0.0071$,
in the case of (a) $I=0.25$, (b) $I=0.5$, (c) $I=2$, and (d) $I=8$.
Impurity concentration $n_{\rm imp}$ is $0.008$.
The insets show that impurity-induced DOS is always 
located at a finite energy for $n_{\rm imp}=0.001$.
}
\label{fig:DOS2}
\end{figure}
%%%%%%%%%%%%%%%%%%%%%%%%%%%%%%%%%

Next, we study the case of $N_\a=1, N_\b=0.5$.
We put $\Delta_\a=0.005$ and $\Delta_\b= -0.0071$ 
by considering the relationship $|\Delta_\a/\Delta_\b|\sim(N_\b/N_\a)^{1/2}$
in the two-band BCS model with repulsive interband interaction,
as explained in \S \ref{sec:T-matrix}.
Figure \ref{fig:DOS2} (a), (b) show the DOS in the superconducting state 
for $n_{\rm imp}=0.008$.
When $I=I'=0.25$, in Fig. \ref{fig:DOS2} (a),
the impurity-induced DOS is reduced
by changing $N_\b$ from $1$ to $0.5$,
by comparing with Fig. \ref{fig:DOS1} (a).
As $x$ decreases from unity, 
the impurity-induced DOS moves to the gap edge.
In the intermediate ($I=0.5$) or unitary ($I\ge2$) regime,
impurity-induced DOS covers the zero energy state 
in the case of $x=1$ and $n_{\rm imp}=0.008$,
as shown in Figs. \ref{fig:DOS2} (b)-(d).
However, finite gap appears around the Fermi level for $x\le0.9$.
In the case of $I=8$, in Fig. \ref{fig:DOS2} (d),
impurity-induced DOS is very large at $x=1$, whereas
it is strongly suppressed for $x\le0.9$.
%The obtained results for $N_\a\ne N_\b$ are similar to those for $N_\a=N_\b$
%given in Figs. \ref{fig:DOS1} (b)-(d).
When the impurity concentration is very low ($n_{\rm imp}\sim 0.001$),
in-gap state deviates from $\w=0$ even if $x=1$
as shown in insets in  Figs. \ref{fig:DOS2} (b)-(d),
since $f_\a+f_\b$ given in eq. (\ref{eqn:local-g})
is non-zero in the case of $N_\a\ne N_\b$ \cite{Bang}.

\section{Discussion}

In the present paper, we studied the impurity effects on the
$s_\pm$-wave superconducting state,
which is expected to be realized in iron oxypnictide superconductors.
There, nonmagnetic impurities can induce both the in-gap bound state and
the reduction in $T_{\rm c}$.
Based on the two-band BCS model, we have found that 
the zero-energy in-gap state emerge under the conditions that
(i) $x\equiv |I'/I|=1$ and (ii) $|I|N_{\a},|I|N_{\b}\gg1$.
Deviating from these conditions,
in-gap state shifts to a finite energy, and disappears eventually.

Here, we discuss the case of unitary scattering:
In iron oxypnictide superconductors, Fe substitution by other elements
(such as Co, Ni, and Zn) will cause the unitary scattering potential.
In this case, the impurity potential is 
diagonal with respect to the $d$-orbital \cite{Senga}.
The impurity potential has off-diagonal elements 
in the band-diagonal representation.
As discussed in Ref. \cite{Senga}, $x\sim 
\langle\sum_d O_{d,\a}(\k)O_{d,\b}(\k')\rangle_{\k\in\a,\k'\in\b}^{\rm FS}$,
where $O_{d,\a}(\k)=\langle d;\k|\a;\k\rangle$ represents the 
transformation matrix between the orbital representation (orbital $d$) 
and the band-diagonal representation (band $\a$).
In iron oxypnictide superconductors, $x\sim 0.5$ 
since the hole-pockets are composed of $d_{xz},d_{yz}$ orbitals of Fe
in the two-Fe unit cell,
whereas half of the electron-pockets are composed of 
$d_{x^2\mbox{-}y^2}$ orbitals \cite{Senga}.
Since the impurity effect is weak except when $x=1$ in the unitary regime 
as shown in Figs. \ref{fig:DOS1} (d) and \ref{fig:DOS2} (d),
Fe substitution by other elements will affect
the superconducting DOS and $T_{\rm c}$ only slightly.

We also discuss the case of Born scattering
due to ``in-plane'' weak random potential or disorder:
As shown in Figs. \ref{fig:DOS1} (a) and (b),
the impurity effect is rather insensitive to $x$.
Therefore, a broad impurity-induced in-gap state 
will emerge in the superconducting DOS, and 
a sizable reduction in $T_{\rm c}$ occurs at the same time.
Born impurity scattering will be also caused by
``off-plane'' impurities like the As substitution by other elements.
In this case, the radius of impurity potential $R$ for Fe sites 
will be about the unit cell length $a$.
Then, the impurity scattering ($\k\rightarrow\k'$) is 
restricted to $|\k-\k'|\lesssim 1/R \sim 1/a$.
Since $|\k-\k'|\approx \pi/a$ in the interband scattering
between electron-pockets and hole-pockets,
$I'$ should be much smaller than $I$.
Therefore, the effect of off-plane impurities on the 
$s_\pm$-wave state will be small 
since the relationship $x\ll1$ is expected to be realized.

In summary, in iron oxypnictide superconductors, Born or intermediate 
in-plane impurities cause prominent impurity effects since the 
$s_\pm$-wave state is violated by the interband scattering.
Only one percent Born impurities with $x\gtrsim0.5$ induce
not only plenty of in-gap DOS, but also sizable reduction in $T_{\rm c}$.
For this reason, relation $1/T_1\propto T^3$ below $T_{\rm c}$
observed in clean LaFeAsO$_{1-x}$F$_{x}$ \cite{Ishida,Grafe} and
in LaFeAsO$_{0.7}$ \cite{Mukuda} samples, which would be almost absent from 
the impurity reduction in $T_{\rm c}$, cannot be explained by 
the present analysis based on the isotropic BCS model.
Thus, anisotropy in the $s_\pm$-wave superconducting gap
might be responsible for the relation $1/T_1\propto T^3$ \cite{Nagai}.
Recently, rapid suppression in $1/T_1$ ($\propto T^{\a}$; $\a>5$)
below $T_{\rm c}$ had been observed in a clean LaFeAsO$_{0.9}$F$_{0.1}$ 
sample with $T_{\rm c}= 28$ K (=intrinsic $T_{\rm c}$) \cite{Sato-T1}.
This result is consistent with the penetration depth 
\cite{Matsuda} and ARPES \cite{ARPES1,ARPES2,ARPES3,ARPES4}, 
and it is naturally explained by the present analysis.
%based on the isotropic BCS model.
Theoretically, in fully gapped $s$-wave superconductor,
the gap function becomes anisotropic due to magnetic fluctuations, 
in a way that the two superconducting gap minima 
are connected by the nesting vector \cite{Boron}.
In iron oxypnictides, 
the degree of anisotropy in the $s_\pm$-wave gap function
is rather sensitive to model parameters
such as the nesting condition \cite{Kuroki,RG}.
The wide variety of behaviors in $1/T_1$ would reflect the large
sample dependence of the gap anisotropy in iron oxypnictide superconductors.

\acknowledgements
We are grateful to M. Sato for enlightening discussions 
on the impurity effect and the gap anisotropy.
We are also grateful to D.S. Hirashima, Y. Matsuda, 
T. Shibauchi, H. Aoki, K. Kuroki, R. Arita, Y. Tanaka, S. Onari and 
Y. ${\bar {\rm O}}$no for useful comments and discussions.
This study has been supported by Grants-in-Aid for Scientific 
Research from MEXT of Japan and from the Japan Society for the 
Promotion of Science, and by JST, TRIP.

%%%%%%%%%%%%%%%%%%%%%%%%
%references
%%%%%%%%%%%%%%%%%%%%%%%%

\end{document}